\documentclass[aps,pra,
 reprint,
 amsmath,amssymb,
 superscriptaddress]{revtex4-2}

\usepackage{graphicx}
\usepackage{dcolumn}
\usepackage{bm}
\usepackage{siunitx}
\usepackage[colorlinks=true,allcolors=blue,bookmarks=false,pdfusetitle]{hyperref}

\usepackage[caption=false]{subfig}
\begin{document}

\title{Deterministic generation of multidimensional photonic cluster states using time-delay feedback}


\author{Yu Shi}
 \email{shiyu@terpmail.umd.edu}
 \affiliation{Department of Electrical and Computer Engineering and Institute for Research in Electronics and Applied Physics, University of Maryland, College Park, Maryland 20742, USA}
\author{Edo Waks}%
 \email{edowaks@umd.edu}
 \affiliation{Department of Electrical and Computer Engineering and Institute for Research in Electronics and Applied Physics, University of Maryland, College Park, Maryland 20742, USA}
 \affiliation{Joint Quantum Institute, University of Maryland, College Park, Maryland 20742, USA}
 \affiliation{Department of Physics, University of Maryland, College Park, Maryland 20742, USA}

\date{\today}

\begin{abstract}
Cluster states are useful in many quantum information processing applications.
In particular, universal measurement-based quantum computation (MBQC) utilizes 2D cluster states \cite{Raussendorf2001}, and topologically fault-tolerant MBQC requires cluster states with three or higher dimensions \cite{Raussendorf2007}.
This work proposes a protocol to deterministically generate multi-dimensional photonic cluster states using a single atom-cavity system and time-delay feedback.
The dimensionality of the cluster state increases linearly with the number of time-delay feedback.
We firstly give a diagrammatic derivation of the tensor network states, which is valuable in simulating matrix product states (MPS) and projected entangled pair states (PEPS) generated from sequential photons.
Our method also provides a simple way to bridge and analyze the experimental imperfections and the logical errors of the generated states.
In this method, we analyze the generated cluster states under realistic experimental conditions and address both one-qubit and two-qubit errors.  Through numerical simulation, we observe an optimal atom-cavity cooperativity for the fidelity of the generated states, which is surprising given the prevailing assumption that higher cooperativity systems are inherently better for photonic applications.
\end{abstract}

\maketitle

\section{Introduction}
Cluster states are highly entangled multi-qubit states that play a significant role in quantum communication and quantum computation \cite{Briegel2001}.
They provide entanglement resources for multi-qubit quantum teleportation and super-dense coding \cite{Muralidharan2008, Liu2014}, in addition to applications in quantum error correction \cite{Schlingemann2002, Bell2014} and quantum metrology \cite{Friis2017, Shettell2020}.
Cluster states also enable universal measurement-based quantum computation \cite{Raussendorf2001, Raussendorf2003}.
However, only cluster states of two dimensions or higher carry such universality, and topologically fault-tolerant cluster state quantum computation requires at least three dimensions \cite{Raussendorf2007}.
Therefore, there is a need to develop new methods to generate multi-dimensional cluster states \cite{Lanyon2013, Pichler2017, Dhand2018, Gimeno-Segovia2019, Larsen2019, Asavanant2019}.

Among various quantum platforms, photons are considered ideal candidates for implementing cluster states because of their pristine coherence properties and precise single-qubit control.
One method to obtain a scalable 1D cluster state is by generating entangled photons from a quantum emitter \cite{Lindner2009, Schwartz2016}.
However, the extension of this approach to higher dimensions requires entangled quantum emitters, which is significantly more challenging \cite{Economou2010, Gimeno-Segovia2019}.
Recently, Pichler and colleagues proposed an alternative approach to achieve 2D cluster states by feeding back the generated photons to the quantum emitter \cite{Pichler2017}.
However, expanding this method to three dimensions or higher, which is essential for fault tolerance \cite{Raussendorf2007}, remains unsolved.

We propose a protocol to generate higher-dimensional photonic cluster states using a single cavity-coupled atom combined with time-delay feedback.
We exploit the atom as an ancilla that applies sequential quantum gates on photons and effectively entangles them.
By looping photons back to the cavity multiple times, we can extend the cluster state's dimensionality to an arbitrary size.
Furthermore, we firstly give a diagrammatic derivation of the tensor network states, which is valuable in simulating matrix product states (MPS) and projected entangled pair states (PEPS) generated from sequential photons.  Our method also provides a simple way to bridge and analyze the experimental imperfections and the logical errors of the generated states.  In this method, we analyze the generated three-dimensional cluster states under realistic experimental conditions and address both one-qubit and two-qubit errors.  We observe a surprising behavior in the fidelity of generated states versus the atom-cavity cooperativity.  It also shows that our approach can generate cluster states with nearly unity fidelity using a chiral coupling atom-cavity system.

\section{\label{sec:two}Protocol}

\begin{figure}[t]
\subfloat[\label{fig:1a}]{}
\subfloat[\label{fig:1b}]{}
\includegraphics[width=3.2in]{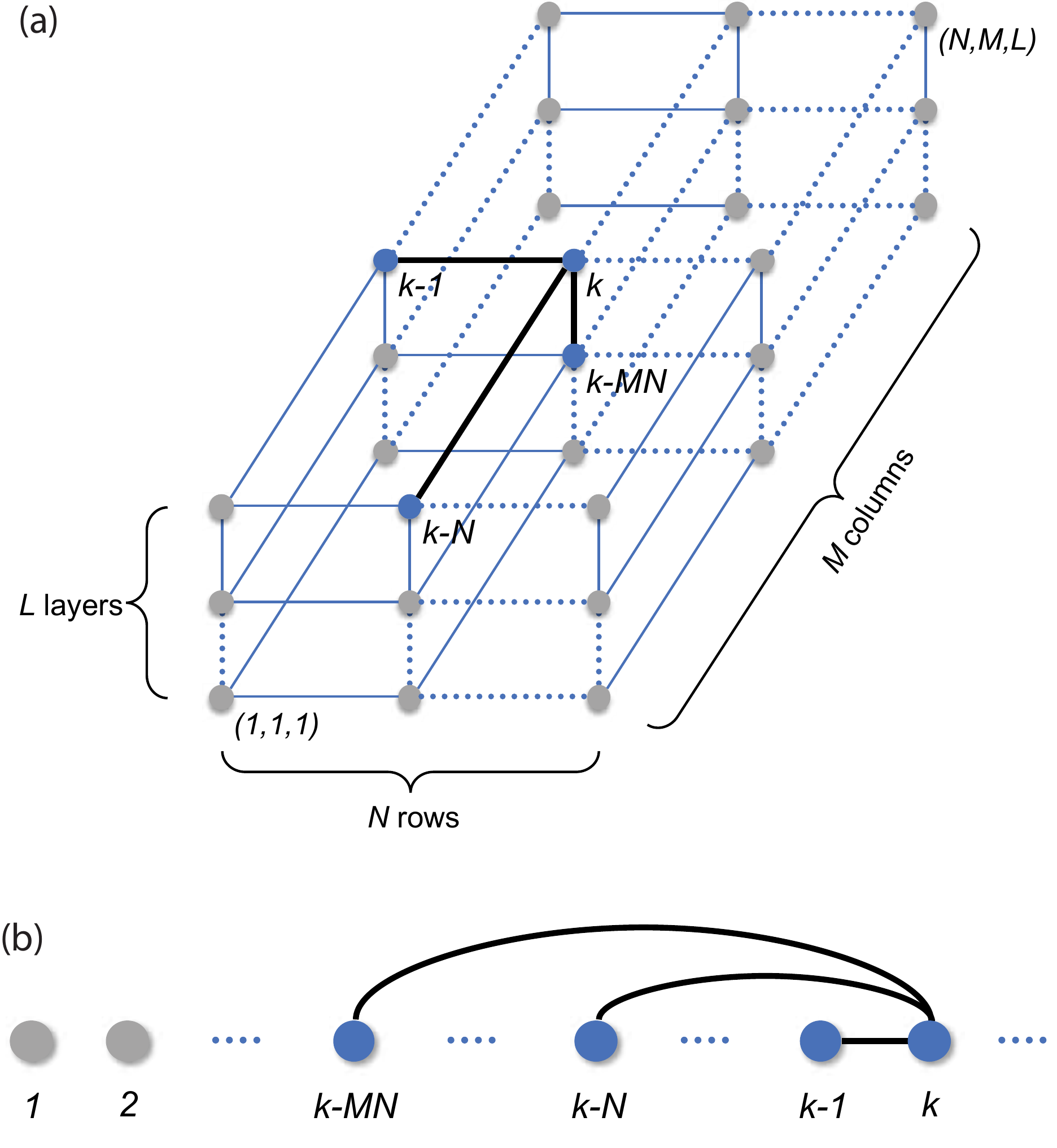}
\caption{\label{fig:1}
(a) The schematic of a 3D cluster state on a cuboid lattice with dimensions $N\times M\times L$.
Each node represents a photon, while the edges between nodes represent entanglement.
The entanglement of photon $k$ with its nearest neighbors $k-1$, $k-N$, and $k-MN$ define the system's unit cell.
(b) By rearranging the 3D lattice into a linear chain, photon $k$ of the unit cell is entangled with one nearest neighbor photon $k-1$, and two non-nearest neighbor photons $k-N$ and $k-MN$.}
\end{figure}

We first consider how to generate 3D cluster states and then generalize to higher dimensions.
A 3D cluster state can be defined on a 3D lattice, in which the nodes represent qubits (i.e., photons in this case), and the edges represent the entanglement between photons.
Figure~\ref{fig:1a} shows a cluster state on a cuboid lattice with dimensions of $N\times M\times L$.
We can give each photon inside the lattice a unique coordinate $\left(n,m,l\right)$ in terms of its row $\left(n\right)$, column $\left(m\right)$, and layer $\left(l\right)$ position, and label it $\left(k\right)$ using the following formula:
\[k=\left(l-1\right)MN+\left(m-1\right)N+n\;.\]
Based on this system, the entanglement of photon $k$ with its nearest neighbors, $k-1$, $k-N$, and $k-MN$, describes the unit cell of the 3D lattice.

Photons are flying qubits, therefore a fixed 3D lattice structure is not convenient for generating a photon-based 3D cluster state.
As an alternative, we can rearrange the lattice as a linear chain.
To form a linear chain from the 3D lattice, we list the photons sequentially by their label $k$ (Figure~\ref{fig:1b}).
The photons then feature a combination of both nearest and non-nearest neighbor entanglement within the chain.
For example, as Figure~\ref{fig:1b} shows, photon $k$ of the unit cell is entangled with one nearest neighbor photon $k-1$, but now photons $k-N$ and $k-MN$ correspond to two non-nearest neighbors.
However, generating this entanglement between nearest and non-nearest neighbors in the linear photon sequence is a challenge, particularly given that photons cannot directly interact with each other.

\begin{figure}[b]
\subfloat[\label{fig:2a}]{}
\subfloat[\label{fig:2b}]{}
\subfloat[\label{fig:2c}]{}
\subfloat[\label{fig:2d}]{}
\includegraphics[width=3.4in]{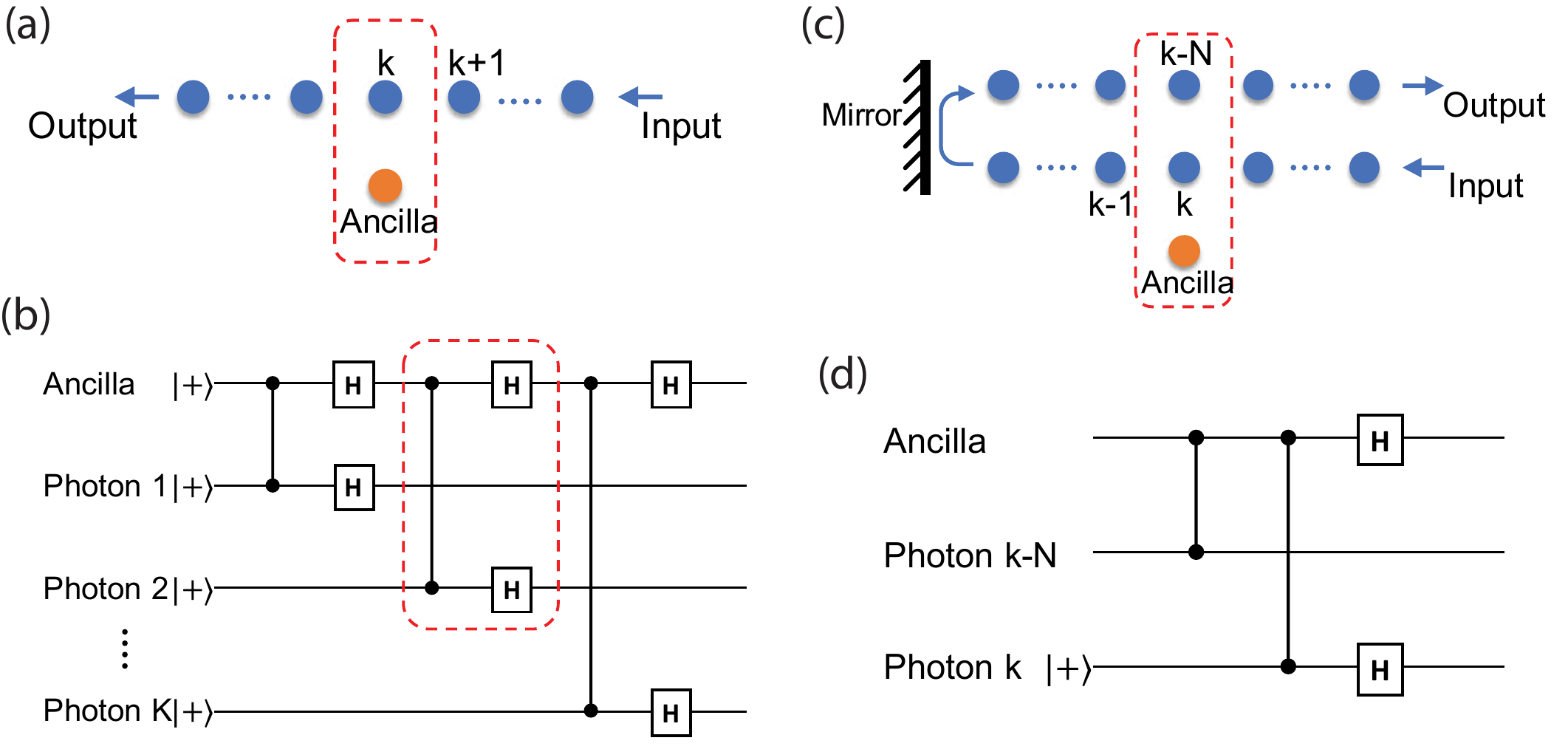}
\caption{\label{fig:2}
(a) The schematic diagram shows a system involving a stationary ancilla and a sequence of photons.
Photons interact with the ancilla one at a time inside the box, corresponding to each process cycle.
(b) The quantum circuit for generating nearest neighbor entanglement between photons.
The vertical lines stand for CPHASE gates, and the squares labeled H stand for Hadamard gates.
(c) The schematic diagram for generating non-nearest neighbor entanglement involving a stationary ancilla, a sequence of photons, and one time-delay photon feedback.
In an actual experiment, photons $k-N$ and $k$ are separated in time and sequentially interact with the ancilla during one cycle.
(d) The quantum circuit for entangling photons $k-N$ and $k$.}
\end{figure}

One way to generate nearest-neighbor entanglement between photons is through sequential interactions with an ancilla (i.e. an atomic qubit).
Figure~\ref{fig:2a} shows how an ancilla can generate a 1D cluster between a sequence of periodically spaced photons with an equal delay time $T_{cycle}$ (one clock cycle).
The photons, represented by blue circles, propagate from right to left and are sufficiently separated to ensure they interact with the ancilla individually and sequentially.
We initialize each photon to the state $|+\rangle=\frac{1}{\sqrt2}\left(|0\rangle+|1\rangle\right)$, where $|0\rangle$ and $|1\rangle$ are the computational basis.
We assume that the ancilla applies a controlled phase flip (CPHASE) gate to each photon, which entangles the photon with the ancilla (later we will describe a specific way to implement this gate using a single atom-cavity system).
We then follow this process with Hadamard gates, as shown in Figure~\ref{fig:2b}.
When the CPHASE and Hadamard gates are applied sequentially in this manner to photons $k$ and $k+1$, they generate an effective CPHASE gate between the two photons (see Appendix~\ref{app:a}).
By iterating this process for each cycle of the experiment for all photons in the linear sequence, we can generate a 1D cluster state, as previously shown by Lindner and Rudolph \cite{Lindner2009}.

Generating non-nearest neighbor entanglement between photons requires time-delay feedback.
Figures~\ref{fig:2c} and~\ref{fig:2d} show how to implement interaction between two photons $k$ and $k-N$.
A mirror reflects photon $k-N$ back to the ancilla with a delay $NT_{cycle}$.
The ancilla subsequently applies a CPHASE gate to photon $k-N$ and then photon $k$.
After this, a Hadamard gate on photon $k$ and the ancilla effectively swap their states, which generates the non-nearest neighbor entanglement between photons $k$ and $k-N$ and prepares the system for the next clock cycle (see Appendix~\ref{app:a}).  Applying this to all photons in the chain generates a 2D cluster state.

\begin{figure}[t]
\subfloat[\label{fig:3a}]{}
\subfloat[\label{fig:3b}]{}
\subfloat[\label{fig:3c}]{}
\subfloat[\label{fig:3d}]{}
\includegraphics[width=3.4in]{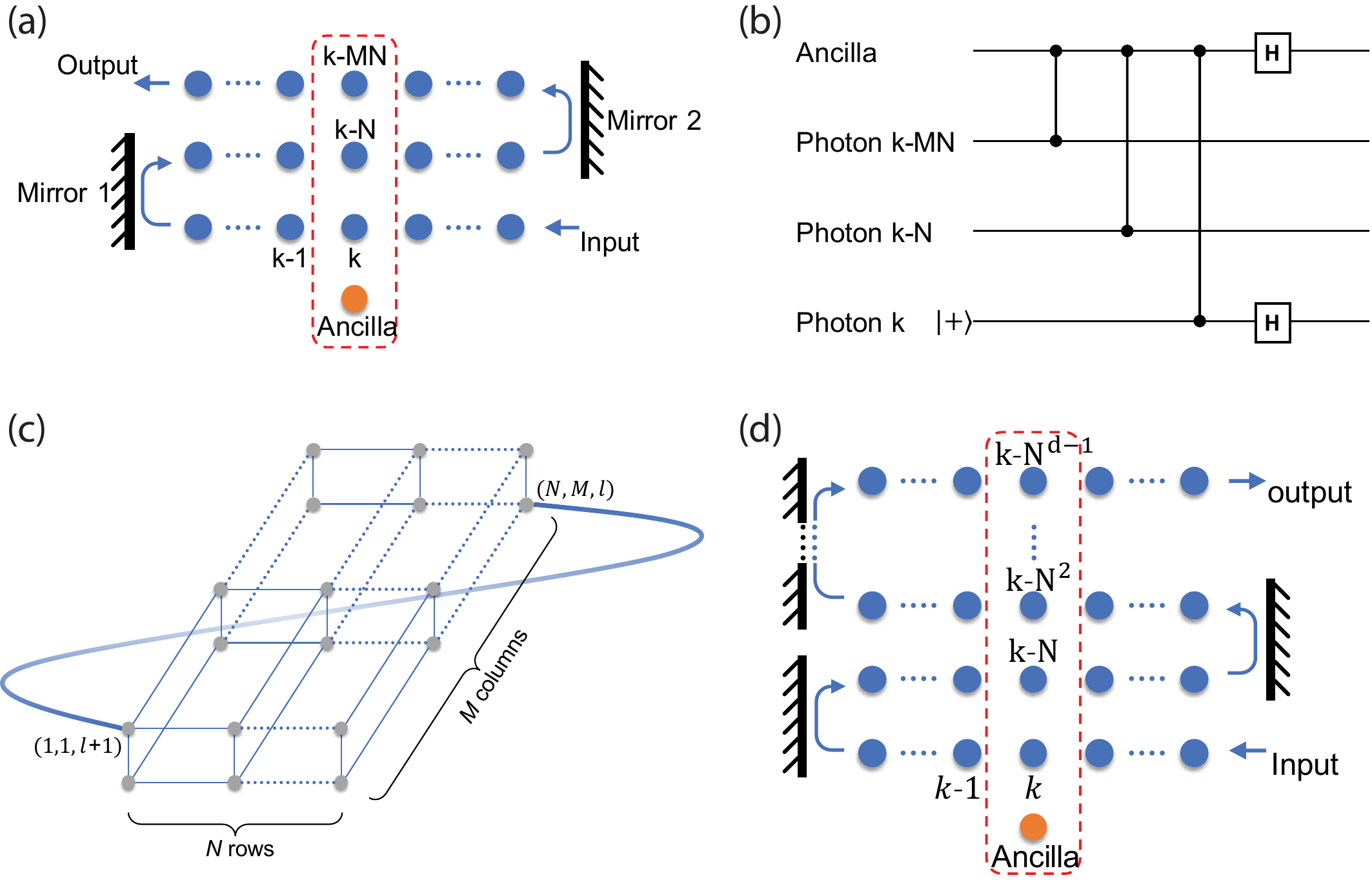}
\caption{\label{fig:3}
(a) The diagram for generating 3D cluster states involving a stationary ancilla, a sequence of injected photons, and two time-delay feedbacks.
In an actual experiment, photons $k-MN$, $k-N$, and $k$ are separated in time and sequentially interact with the ancilla during one cycle.
(b) The quantum circuit for entangling photons $k-MN$, $k-N$, and $k$.
(c) An example of the helical entanglement of a 3D cluster state, where boundary photons $\left(N,M,l\right)$ and $\left(1,1,l+1\right)$ are entangled.
(d) The schematic diagram for generating $d$-dimensional cluster states using $(d-1)$ time-delay feedbacks.}
\end{figure}

Now that we have a way to generate both nearest and non-nearest neighbor interactions, we can combine them to form multi-dimensional cluster states.
Figure~\ref{fig:3a} shows how to do this for the case of a 3D cluster state, which requires two non-nearest neighbor interactions, therefore two time-delay feedbacks.
The first mirror delays a photon for $NT_{cycle}$ and the second delays for $\left(MN-N\right)T_{cycle}$, which allows the ancilla to apply a CPHASE gate to photons $k-MN$, $k-N$, and $k$ sequentially.
The protocol is actually agnostic to the order with which the CPHASE gates are applied, but we assume this particular order for concreteness.
After the CPHASE gates, we apply a Hadamard gate on photon $k$ and the ancilla, which realizes the quantum circuit in Figure~\ref{fig:3b}.
Iterating this circuit sequentially over each cycle for all the photons yields a 3D cluster state.

The procedure outlined above generates a cluster state that is slightly different from the one illustrated in Figure~\ref{fig:1a}.
The interior of the cluster state is identical, but the edge produces a more complicated boundary condition as shown in Figure~\ref{fig:3c}.
Photon $(N,M,l)$, which is on the lattice boundary, is also entangled with photon $(1,1,l+1)$, because these two photons are next to each other in the rearranged linear chain.
Therefore, the resulting cluster state features helical connectivity, as was the case for the 2D cluster state proposed by Pichler et al. \cite{Pichler2017}, but now extended to a higher dimension.
Note also that the elementary cell in the fault-tolerant cluster states is not a primitive cubic as shown in Figure~\ref{fig:1a} \cite{Raussendorf2007}, but we can obtain it by measuring and eliminating the extra qubits after the generation process.

The extension to even higher dimensions follows analogously.
Each dimension requires an additional non-nearest neighbor connection, which necessitates another time-delay feedback.
As illustrated in Figure~\ref{fig:3d} for generating $d$-dimensional cluster states, we use $(d-1)$ mirrors to implement $(d-1)$ time-delay feedbacks.
Therefore, the protocol is efficient in the sense that increasing the dimensions results in only a linear increase in the number of time-delay feedback.

\section{Physical implementation}
\begin{figure}[b]
\subfloat[\label{fig:4a}]{}
\subfloat[\label{fig:4b}]{}
\includegraphics[width=3.2in]{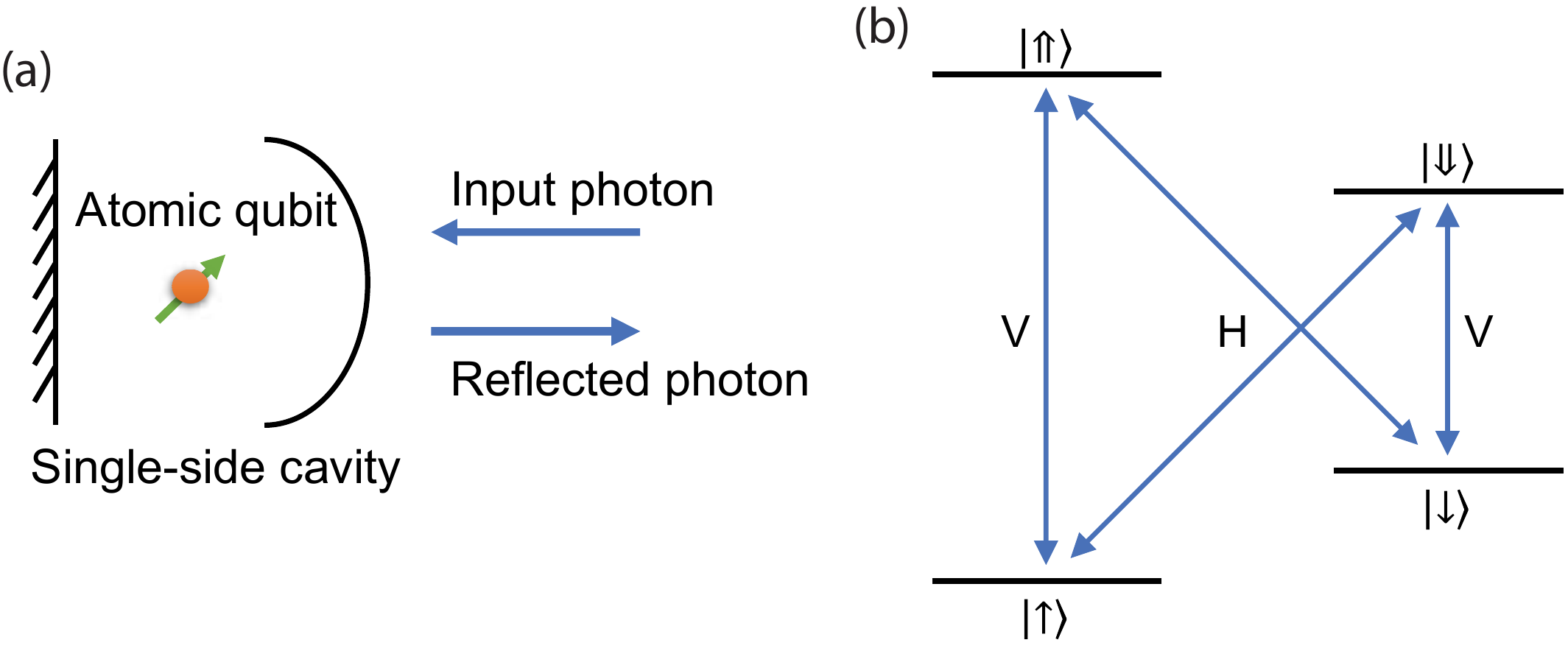}
\caption{\label{fig:4}
(a) The schematic implementation of the CPHASE gate achieved by reflecting a photon from the single-side cavity containing an atomic qubit coupled single-side cavity.
(b) The atomic-level structure of the cavity-coupled atomic qubit, which features four optical transitions. V(H) means the vertical(horizontal) polarization of light.}
\end{figure}

The previous section provides an abstract outline of the protocol of generating multi-dimensional cluster states.
Here we describe a method to implement this concept based on photonics and cavity quantum electrodynamics.
We consider the specific case where the ancilla is composed of an atomic spin qubit coupled to an optical single-sided cavity, as illustrated in Figure~\ref{fig:4a}.
These systems are already known to generate CPHASE gates between the atomic and photonic qubits using cavity-mediated interactions, as proposed theoretically \cite{Duan2004} and realized experimentally using numerous atomic systems, including quantum dots \cite{Kim2013, Sun2016, Sun2018, Luo2019}, atoms \cite{Reiserer2014, Hacker2016}, and color centers in diamond \cite{Evans2018}.

We consider an atom in a cavity where an external magnetic field is applied in a direction parallel to its cross-plane.
Figure~\ref{fig:4b} illustrates this specific atomic-level structure.
The atomic spin qubit possesses two ground states $(\left|\uparrow\right\rangle, \left|\downarrow\right\rangle)$ and two excited states $(\left|\Uparrow\right\rangle, \left|\Downarrow\right\rangle)$, where the quantization axis is along the direction of the magnetic field.
The atom features four optical transitions.
The vertical transitions couple to a linearly polarized photon whose polarization is parallel to the quantization axis; the cross transitions couple to a linearly polarized photon whose polarization is perpendicular to the quantization axis \cite{Press2008}.
We assume the cavity only supports one linearly polarized mode that enables the vertical transitions.
We define this polarization direction as the vertical (V) polarization and its orthogonal direction as the horizontal (H).

When the probe beam (external photon) is in the horizontal polarization, it will reflect off a mirror and acquire no phase shift.
When the probe beam is in the vertical polarization, the reflection coefficient of the cavity is given by \cite{Walls2008} (see Appendix~\ref{app:b})
\begin{equation}
    r_{\uparrow,\downarrow}=\frac{g^2-\frac{\gamma}{2}\left[i\Delta_{\uparrow,\downarrow}+\frac{\kappa}{2}\right]}{g^2+\frac{\gamma}{2}\left[i\Delta_{\uparrow,\downarrow}+\frac{\kappa}{2}\right]}\;,
\end{equation}
where $\Delta_{\uparrow,\downarrow}$ is the frequency detuning between the photon and atomic transition; $g$, $\kappa$, and $\gamma$ are the atom-cavity coupling strength, atom dipole decay rate, and cavity decay rate, respectively.  
We assume the spin-up transition is on-resonant with the frequency of the photon $\left(\Delta_\uparrow=0\right)$, therefore the spin-down transition frequency detunes $\Delta_\downarrow$ from the photon frequency by the Zeeman splitting.
When the spin is in state $\left|\uparrow\right\rangle$, the reflection coefficient $r_\uparrow=\frac{C_\uparrow-1}{C_\uparrow+1}$ where we define $C_\uparrow=\frac{4g^2}{\gamma\kappa}$ as the on-resonant atom-cavity cooperativity.
When the spin is in state $\left|\downarrow\right\rangle$, the reflection coefficient $r_\downarrow=\frac{C_\downarrow-1}{C_\downarrow+1}$ where we define $C_\downarrow=\frac{4g^2}{\gamma\left(\kappa+i2\Delta_\downarrow\right)}$ as the off-resonant cooperativity.  
In the limit of high on-resonant cooperativity $\left(C_\uparrow\gg1\right)$ and large detuning $\left(\Delta_\downarrow\gg\frac{\kappa C_\uparrow}{2}\right)$, the two reflection coefficients approach $r_\uparrow=1$ and $r_\downarrow=-1$, repectively.
We define the state of the incident photon as $\left|H\right\rangle\equiv\left|0\right\rangle_p$ for a horizontally polarized photon and $\left|V\right\rangle\equiv\left|1\right\rangle_p$ for a vertically polarized photon.
We similarly label the spin states as  $\left|\uparrow\right\rangle\equiv\left|0\right\rangle_A$ and $\left|\downarrow\right\rangle\equiv\left|1\right\rangle_A$.
The state of the total system, denoted as $\left|\psi\right\rangle=\left|x\right\rangle_p\ \otimes\left|y\right\rangle_A$ where $x$ and $y$ are the qubit states of the photon and spin respectively, transforms according to the following operator
\begin{equation}
U_{RF}=
    \begin{pmatrix}
    1 & 0 & 0 & 0\\
    0 & 1 & 0 & 0\\
    0 & 0 & r_{\uparrow} & 0\\
    0 & 0 & 0 & r_{\downarrow}
    \end{pmatrix}\;.
\label{eq:2}
\end{equation}
The reflection of the photon from the cavity applies a CPHASE gate between the photon and atomic qubit.

\begin{figure}[t]
\subfloat[\label{fig:5a}]{}
\subfloat[\label{fig:5b}]{}
\subfloat[\label{fig:5c}]{}
\includegraphics[width=3.4in]{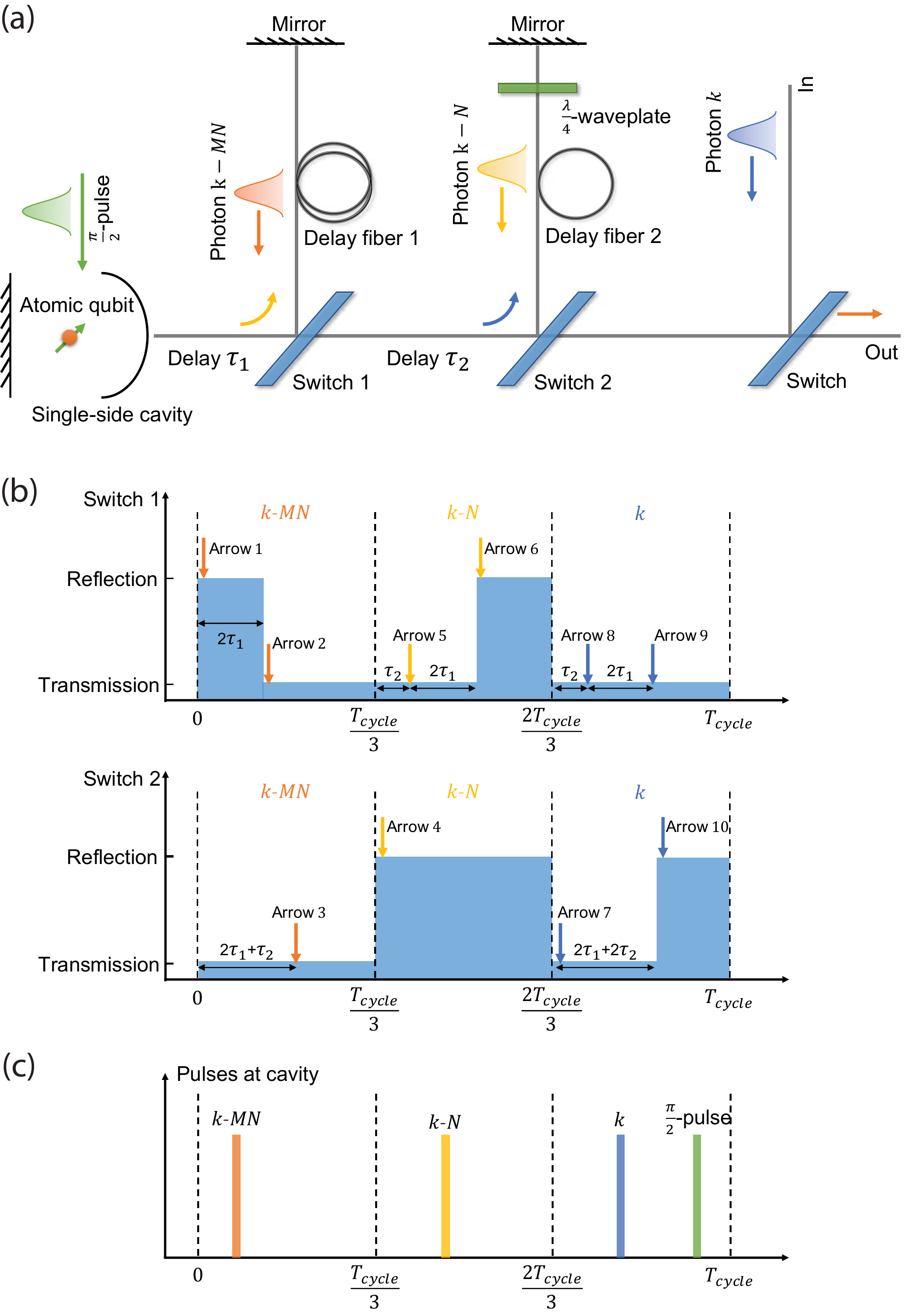}
\caption{\label{fig:5}
(a) The physical setup for generating a 3D cluster state.
At the $k^{th}$ cycle, photons $k-MN$, $k-N$, and $k$ sequentially reflect off the cavity.
(b) Schematic diagrams showing how the two switchable mirrors flip their states between reflective and transmissive during a photon injection period $\left(T_{cycle}\right)$.
The arrows indicate the time points when photons reach the switches.
(c) A diagram showing the sequence of pulses reaching the cavity during $T_{cycle}$.}
\end{figure}

Figure~\ref{fig:5a} shows how we can physically implement the time-delay feedback required to generate a 3D cluster state.
The two time-delay feedback loops consist of two delay lines (delay line 1 and 2) and two fast optical switches (switch 1 and 2) that can be rapidly changed to either reflect or transmit photons.
We insert a $\frac{\lambda}{4}$-waveplate into delay line 2 to rotate the polarization of the photons.
We also insert a switch in the input/output port of the system to direct the input photons toward the cavity and couple the photons leaving the cavity with the output port.
We denote the time delay between sequentially injected photons as $T_{cycle}$.
To generate a 3D cluster state with a dimension of $N\times M\times L$, we set the delay of line 1 as $\left(MN-N-\frac{1}{3}\right)T_{cycle}$ and that of line 2 as $\left(N-\frac{1}{3}\right)T_{cycle}$.
We define the time delay between switch 1 and the cavity as $\tau_1$, the delay between switch 1 and switch 2 as $\tau_2$, in which $\tau_1,\tau_2\ll T_{cycle}$.

Figure~\ref{fig:5b} shows how we set the reflection and transmission states of switch 1 and switch 2 as a function of time relative to the beginning of the $k^{th}$ cycle, denoted as $t=0$.
We divide the clock cycle $T_{cycle}$ into three sections corresponding to the propagating periods of photons $k-MN$, $k-N$, and $k$, respectively.
The arrows denote the timings when a photon arrives at each switch, either in the forward direction to the cavity or in the backward direction after reflecting from the cavity.
The timings are selected such that photon $k-MN$ first interacts with the cavity and leaves the system, then photon $k-N$ reflects off of the cavity and is injected into delay line 1, then photon $k$ reflects off of the cavity and is injected into delay line 2.
These steps implement the CPHASE gates between the photons and the ancilla shown in the quantum circuit in Figure~\ref{fig:3b}.
Then, in delay line 2, photon $k$ passes through the $\frac{\lambda}{4}$-waveplate, reflects off a mirror and passes through the waveplate again, which realizes an effective $\frac{\pi}{2}$ rotation, thus performing a Hadamard gate on the photon polarization \cite{OBrien2007}.
Finally, we perform a Hadamard gate on the atomic qubit using an ultrafast optical Raman control pulse of $\frac{\pi}{2}$ rotation \cite{Press2008, Carter2013}, as shown in the timing diagram in Figure~\ref{fig:5c}, in which the $\frac{\pi}{2}$-pulse arrives at the cavity after the sequential reflections of photons $k-MN$, $k-N$, and $k$ in the cycle.

For the first few photons $1$ through $MN$, either delay line 1 or delay line 2 may not contain any photons.
However, the procedure still works and does not need to be modified in this initial phase.
For photons $1$ through $N$, both delay lines are empty, so these photons will reflect off the cavity sequentially and build an initial 1D cluster state.
Photons $N+1$ through $MN$ will then entangle with the returning photons from delay line 2, but no photon will leave from delay line 1.
Therefore, these photons will build the initial 2D cluster state that forms the first layer of the 3D cluster state.
The remaining photons will then build the additional layers required to form the 3D lattice.
Note also that the number of injected photons increases by one on-demand such that its last layer can be partially filled.

\begin{figure}[b]
\subfloat[\label{fig:6a}]{}
\subfloat[\label{fig:6b}]{}
\includegraphics[width=3.4in]{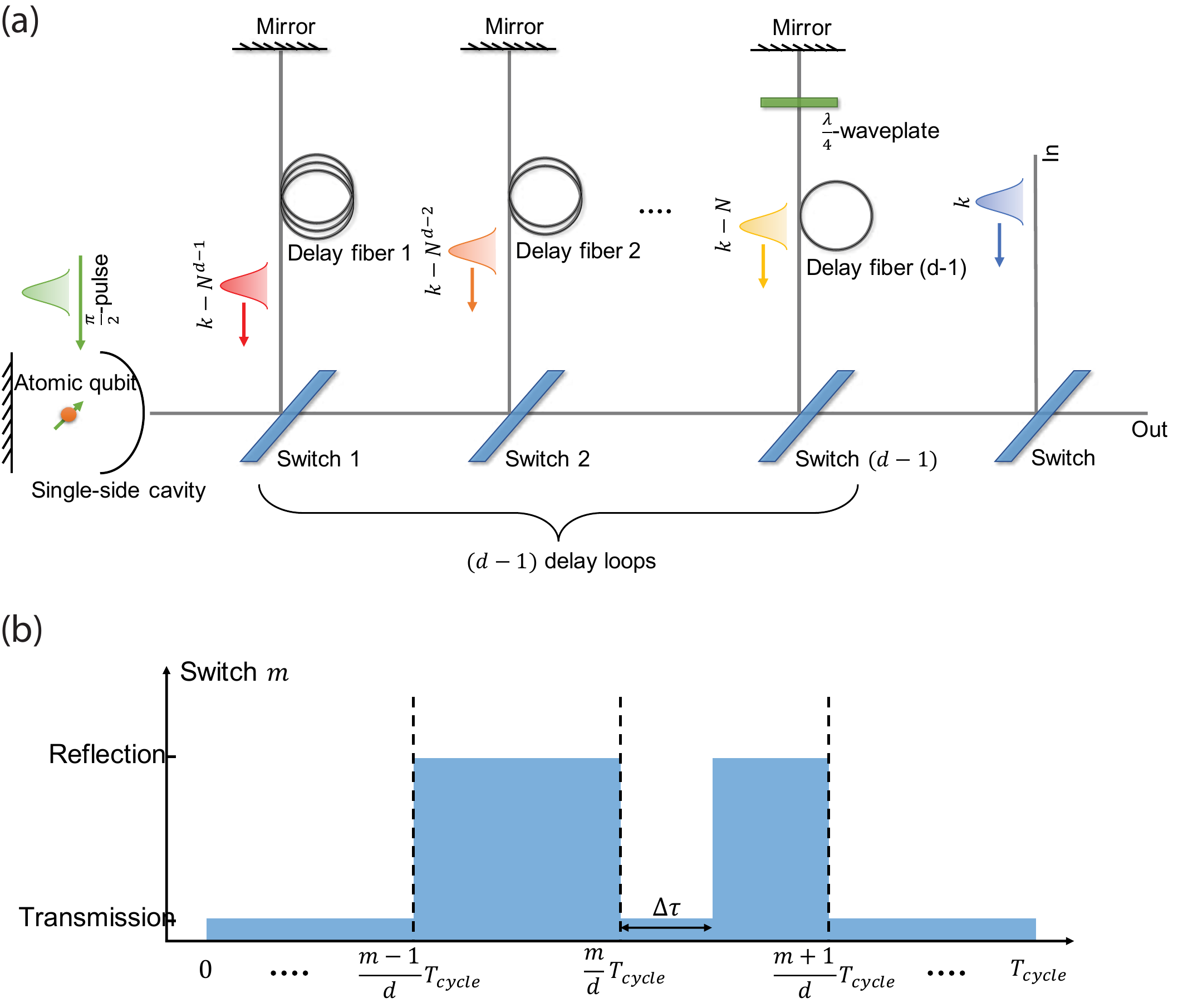}
\caption{\label{fig:6}
(a) The physical setup for generating a $d$-dimensional cluster state.
(b) The schematic diagram for the state of switch $m$.}
\end{figure}

To extend to higher dimensions, we can add more time-delay feedback loops, as shown in Figure~\ref{fig:6a}.
The system consists of $(d-1)$ time-delay feedback loops.
We set the delay time of line $m$ as $\left(\prod_{l=1}^{m}N_l-\prod_{l=1}^{m-1}N_l-\frac{1}{d}\right)T_{cycle}$, where $N_l$ is the length of the $l^{th}$ dimension of the cluster state.
Figure~\ref{fig:6b} shows the timing diagram for switch $m$ during a full cycle of duration $T_{cycle}$.
The timing diagram is an extension of Figure~\ref{fig:5b} but now compatible with more feedback loops.

\section{Analysis}
To analyze the performance of the protocol, we utilize a tensor network formalism \cite{Orus2014}.
A tensor network formalism can efficiently represent a cluster state by several connected tensors equal to the number of photons.
Additionally, the tensor network allows us to calculate the fidelity of the practically implemented cluster state compared to its theoretical prediction by drawing diagrams that make the calculation more visual and straightforward than the traditional linear algebra method.
This section first provides a diagrammatic derivation of the tensor network representation of the multi-dimensional cluster state.  Our derivation is not limited to analyzing the cluster state but is valuable on a broader scenario of simulating matrix product states (MPS) \cite{Vidal2004} and projected entangled pair states (PEPS) \cite{Verstraete2004} generated by sequential photons \cite{Schon2005, Schon2007, Xu2018, Lubasch2018, Gimeno-Segovia2019}.

\begin{figure}[t]
\subfloat[\label{fig:7a}]{}
\subfloat[\label{fig:7b}]{}
\subfloat[\label{fig:7c}]{}
\includegraphics[width=3.2in]{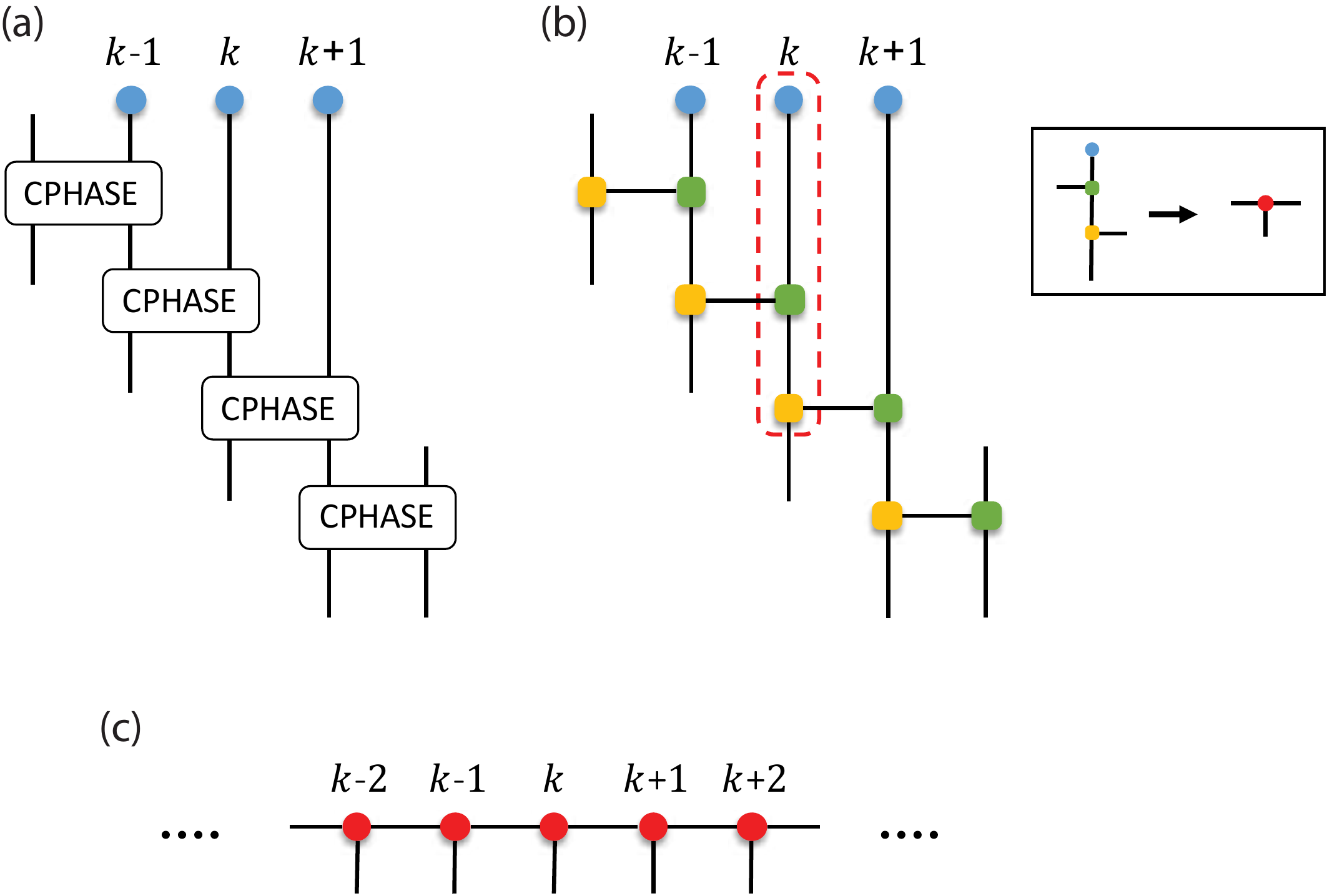}
\caption{\label{fig:7}
(a) The tensor network diagram that produces a 1D cluster state.
(b) The diagram after decomposing the CPHASE tensors in (a) into two rank-3 tensors represented by the yellow and green squares, which are then combined with the circle tensor to form a new rank-3 tensor.
(c) Applying the tensor contraction process shown in (b) for all photons allows us to derive the simplified tensor network representing a 1D cluster state.}
\end{figure}

Figure~\ref{fig:7} illustrates the tensor network representation for a 1D cluster state, which we consider first before extending to higher dimensions.
In 1D cluster states, photon $k$ is connected to two photons, $k-1$ and $k+1$.
Figure~\ref{fig:7a} shows the tensor network diagram where the photons' states and the two-qubit gates each have a tensor representation.
The blue circles are rank-1 tensors representing initially unentangled photons in state $|+\rangle$.
The boxes labeled CPHASE are rank-4 tensors representing CPHASE gates.
The lines emerging from the blue circles and boxes represent the indices of the tensors.
Each rectangle box has four lines, two pointing upward and two downward.
The upward lines represent the two input qubits, and the downward lines represent the two output qubits.
The lines connecting the circles and boxes represent the application of CPHASE gates between two photons.
The uncontracted downward lines represent the qubits for photons of the generated cluster state.

In canonical tensor network state representations such as MPS and PEPS, each tensor represents one qubit, which may have multiple contracted indices but only one uncontracted index.
However, the diagram in Figure~\ref{fig:7a} features more tensors than qubits in the system, which needs to be simplified.
We first decompose the CPHASE tensor into two rank-3 tensors using singular value decomposition (see Appendix~\ref{app:c}).
Figure~\ref{fig:7b} shows the diagram after the decomposition.
The yellow and green squares represent the two tensors decomposed from the CPHASE tensors, consisting of two vertical lines and one horizontal line.
Unlike the CPHASE tensors, the yellow and green tensors operate on one photon.
We can then contract tensors for a photon along its vertical line (see the dashed box in Figure~\ref{fig:7b}), which allows us to obtain a new rank-3 tensor (see the solid box in Figure~\ref{fig:7b}).
This rank-3 tensor has one uncontracted index representing the qubit basis and two contracted indices representing entanglement with its neighbors.
Applying this tensor contraction process for all photons results in the canonical tensor network diagram of the 1D cluster state, as shown in Figure~\ref{fig:7c}.

\begin{figure}[t]
\subfloat[\label{fig:8a}]{}
\subfloat[\label{fig:8b}]{}
\subfloat[\label{fig:8c}]{}
\includegraphics[width=3.4in]{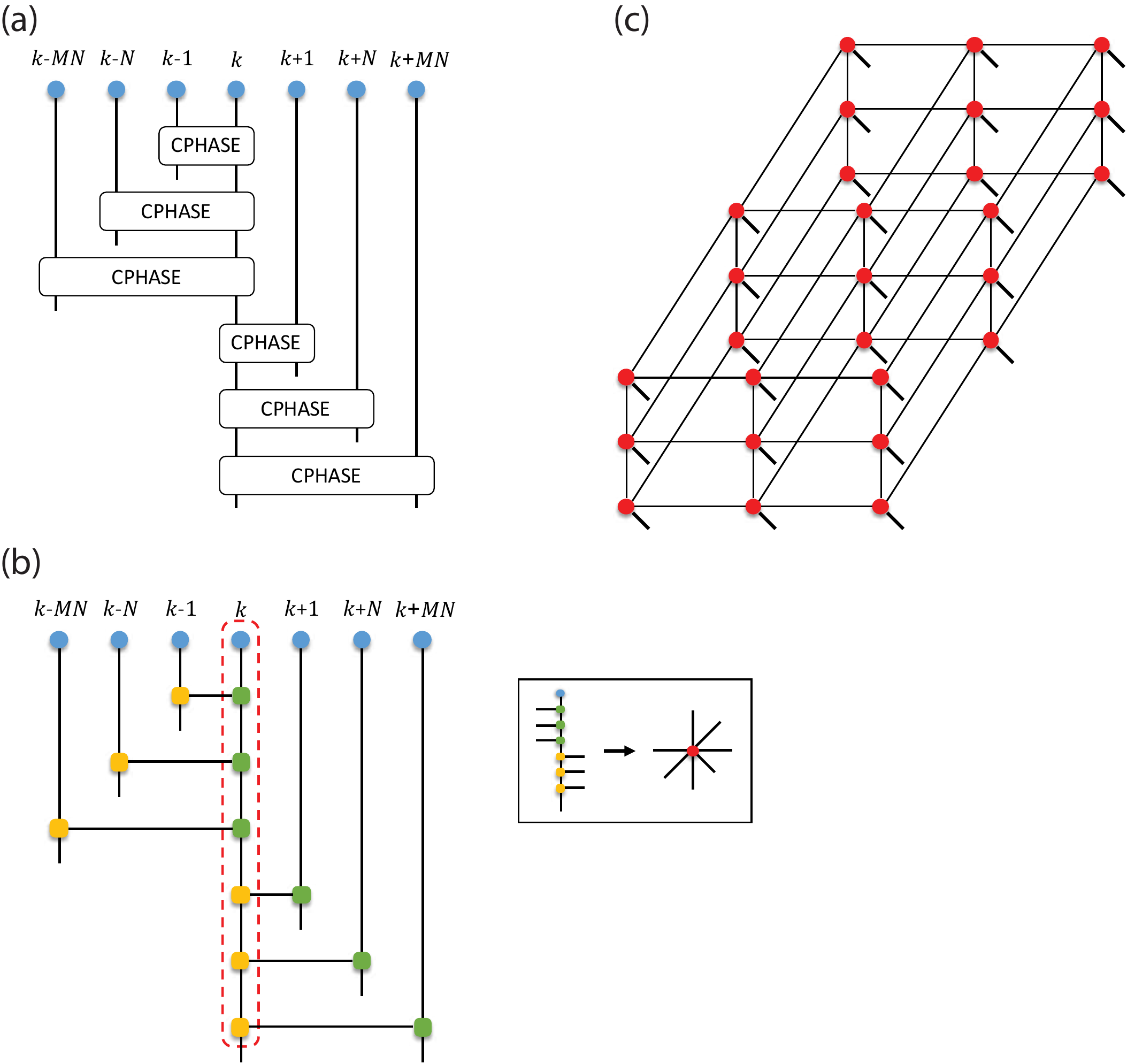}
\caption{\label{fig:8}
(a) The tensor network diagram that entangles photon $k$ with its neighbors in the 3D cluster state.
(b) The diagram after decomposing the CPHASE tensors in (a) into two rank-3 tensors represented by yellow and green squares, which can then be combined with the circle tensor into a rank-7 tensor.
(c) Applying the tensor contraction process shown in (b) for all photons allows us to derive the simplified tensor network representing a 3D cluster state.}
\end{figure}

Figure~\ref{fig:8} shows how we extend the tensor network representation to a 3D cluster state.
In three dimensions, a photon connects to six photons.
Figure~\ref{fig:8a} shows how we build those connections in our protocol using two time-delay feedbacks.
To simplify the tensor network, we again decompose the CPHASE tensors into two rank-3 tensors each (green and yellow squares) to obtain the new diagram, as shown in Figure~\ref{fig:8b}.
Contracting the tensors along the vertical line for a photon then results in a rank-7 tensor.
This tensor consists of one uncontracted index representing the basis of a qubit and six contracted indices representing entanglement with its neighbors.
Applying this tensor contraction process for all photons results in the tensor network diagram of the 3D cluster state, as shown in Figure~\ref{fig:8c}.

We can easily extend this derivation to higher $d$-dimensional cluster states.
For each additional dimension, we add two more square tensors to the contraction sequence.
Finally, we obtain a rank $2d+1$ tensor that features a single uncontracted index representing the qubit basis and $2d$ contracted indices representing entanglement with its neighbors.
The resulting tensor network representation for the $d$-dimensional cluster state consists of these contracted tensors in a $d$-dimensional superlattice.

The previous derivation assumes ideal CPHASE gates, a bad assumption in a realistic system.
In the real physical implementation, we carry out the spin-photon gate by a reflection operation $U_{RF}$ as shown in Eq.~\ref{eq:2}.
Because of finite cooperativity and detuning, this reflection operation will lead to an imperfect CPHASE gate.
However, we can also incorporate $U_{RF}$ in the tensor network formalism.
In Appendix~\ref{app:d}, we show how to transfer the imperfection of $U_{RF}$ to the photon-photon CPHASE gate.
The imperfect CPHASE gate can still be represented by a rank-4 tensor that can be converted to a canonical representation using the decomposition illustrated in Figure~\ref{fig:8}.

Having incorporated imperfect CPHASE gates into the tensor network formalism, we can calculate the fidelity $\left(\mathcal{F}_0\right)$, which serves as the merit for the protocol's performance.
We define this fidelity as
\begin{equation}
\mathcal{F}_0=\frac{\left|\left\langle\Psi_C\middle|\Phi_C\right\rangle\right|^2}{\left\langle\Phi_C\middle|\Phi_C\right\rangle}\;,
\end{equation}
where $\left|\Psi_C\right\rangle$ is the ideal cluster state, and $\left|\Phi_C\right\rangle$ is the unnormalized output state generated by imperfect CPHASE gates.
Using the decomposition described in the previous paragraph, we can represent these two states by tensor networks and calculate the inner product of the two states by numerically contracting the networks.
To perform this calculation, we used the ITensor library software \cite{Fishman2020}.

The previous definition of fidelity does not account for spin decoherence, which arises from the spin's interaction with its nearby spins \cite{Khaetskii2002, Maze2008}.
We can model this decoherence process as a quantum phase-flip channel that randomly flips the spin’s phase with a probability of $\frac{T_{cycle}}{T_2}$ in one cycle, where $T_2$ is the characteristic spin dephasing time \cite{Michael2010}.
This phase-flip error is equivalent to flipping the phase of the next injected photon which transforms the cluster state into another state that resides in an orthogonal subspace \cite{Lindner2009, Raussendorf2003}.
Thus the spin decoherence decreases the fidelity of the cluster state by a factor of $\left(1-\frac{T_{cycle}}{T_2}\right)$ in one cycle.

In addition to decoherence, our fidelity definition must also consider the error of imperfect spin rotation.
Similarly, we can model the imperfect spin rotation as a quantum depolarization channel on the next injected photon which occurs at the rate $p$ and decreases the fidelity by $\left(1-\frac{p}{2}\right)$ \cite{Michael2010}.
By combining the spin dephasing and imperfect spin rotation, the total spin errors decrease the fidelity by a factor of $\alpha=\left(1-\frac{p}{2}\right)\left(1-\frac{T_{cycle}}{T_2}\right)$ in one photon injection period.
Therefore, we obtain a revised fidelity $(\mathcal{F}_1)$ of
\begin{equation}
\mathcal{F}_1=\alpha^N\cdot \mathcal{F}_0\;,
\end{equation}
where $N$ is the number of photons in the cluster state.

Photon loss is another source of error that decreases the generation rate of cluster states. However, this error can be easily detected by comparing the number of photons that enter and leave the system.
To estimate photon loss, we define the probability of collecting all the photons after one run of the protocol as the success probability $\left(\mathcal{P}\right)$, which can be calculated by
\begin{equation}
    \mathcal{P}=\eta^N\cdot\left\langle\Phi_C\middle|\Phi_C\right\rangle\;,
\end{equation}
where $\eta$ is the detection efficiency that accounts for state-independent loss, and $\left|\Phi_C\right\rangle$ is the unnormalized output state accounting for the loss from finite cooperativity.

\section{Simulation}
To analyze the protocol's performance under realistic experimental conditions, we consider the physical implementation of a spin qubit based on an electron charged quantum dot in a nano-cavity \cite{Hennessy2007, Carter2013, Schaibley2013, Sun2016}.
This system implements the spin-photon CPHASE gate modeled by $U_{RF}$ in Eq.~\ref{eq:2}.
The two spin-dependent reflection coefficients are calculated as $r_{\uparrow,\downarrow}=\frac{C_{\uparrow,\downarrow}-1}{C_{\uparrow,\downarrow}+1}$, where $C_\uparrow=\frac{4g^2}{\gamma\kappa}$ is the on-resonant atom-cavity cooperativity, and $C_\downarrow=\frac{4g^2}{\gamma(\kappa+i2\Delta_\downarrow)}$ is the off-resonant cooperativity.
$g$ is the quantum dot-cavity coupling strength, $\kappa$ is the trion state dipole decay rate, $\gamma$ is the cavity decay rate, and $\Delta_\downarrow$ is the detuning frequency between the spin-down state transition and the cavity mode.
The detuning is determined by an applied magnetic field $(B)$ as $\Delta_\downarrow=\left(g_e+g_h\right)\mu_BB/\hbar$, where $g_e$ and $g_h$ are Lande factors for the electron and hole, respectively; $\mu_B$ is the Bohr magneton.
Based on previous experimental measurements, we select the following values for the system's parameters: $g_e=0.43$, $g_h=0.21$, $g/2\pi=\SI{10}{\giga\hertz}$, $\kappa/2\pi=\SI{0.3}{\giga\hertz}$, $\gamma/2\pi=\SI{40}{\giga\hertz}$ \cite{Sun2016}.
We also choose a spin rotation fidelity of $98\%$ and a spin characteristic dephasing time of $T_2=\SI{2}{\micro\second}$ \cite{Press2010, Koppens2008}.
With these parameters, we can calculate the fidelity of the generated state under realistic experimental conditions.

\begin{figure}[b]
\subfloat[\label{fig:9a}]{}
\subfloat[\label{fig:9b}]{}
\subfloat[\label{fig:9c}]{}
\includegraphics[width=3.4in]{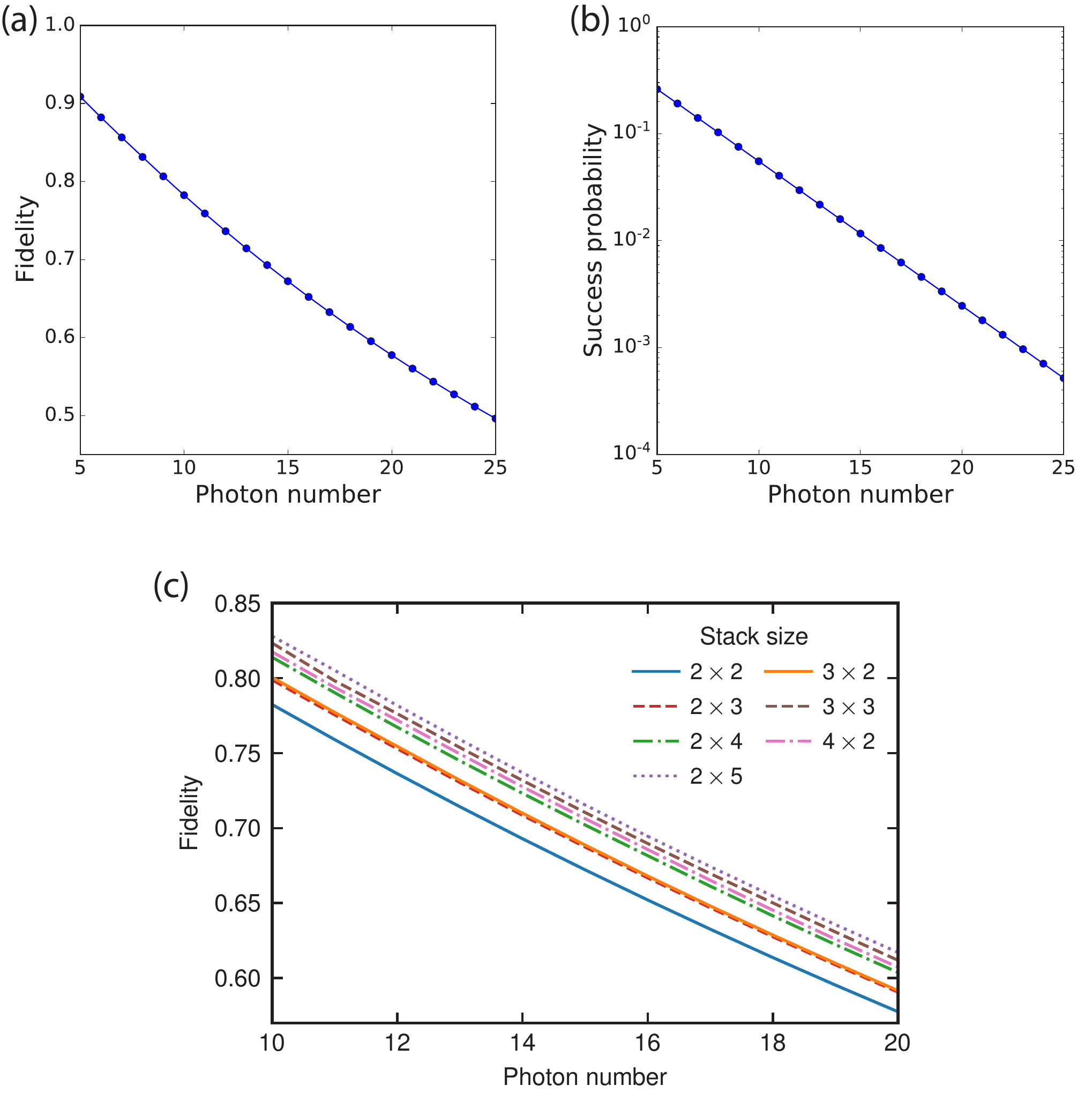}
\caption{\label{fig:9}
(a) The fidelity as a function of the photon number for a 3D cluster state with a $2\times2$ stack, which forms the $N\times M$ layer for a 3D lattice with dimensions of $N\times M\times L$.
(b) The success probability as a function of the photon number for the same 3D cluster state.
(c) Fidelity as a function of photon number for several different stack sizes.}
\end{figure}

We begin with the simplest 3D cluster state with a dimension of $2\times2\times L$, where $L$ can extend to any number because evaluating the contraction of two 3D tensor network states with dimensions of $N\times M\times L$ only requires storing an $N\times M$ tensor network in memory \cite{Orus2014}.
Figure~\ref{fig:9a} shows the fidelity as a function of the cluster state’s photon number with the applied magnetic field set to $B=\SI{12}{\tesla}$.
We find that fidelity decreases with increasing photons, mainly due to the imperfect spin-photon CPHASE gates.
By fitting the data with an exponential function, we see a scaling factor of the fidelity $\left(\beta\right)$ of $0.97$ per photon of the cluster states $\left(\mathcal{F}_1\sim \beta^K\right)$, where $K$ is the photon number of the cluster state.
Figure~\ref{fig:9b} plots the success probability as a function of the photon number for the same cluster state as in Figure~\ref{fig:9a}, in which we assume a detection efficiency of $\eta=0.8$ \cite{Bhaskar2020}.
Using this result, we can estimate the generation rate of the cluster state.
For example, the success probability of an 8-photon cluster state is $0.103$.
If we set the photon injection period as $T_{cycle}=\SI{5}{\nano\second}$, the system can output one state every $\SI{40}{\ns}$, which successfully generates the cluster state at a rate of $2.6\times{10}^6$ per second.

Having shown the fidelity of the simplest 3D cluster states in a $2\times2\times L$ lattice that features a $2\times2$ stack, we investigate the fidelity for a larger stack size.
Figure~\ref{fig:9c} shows the fidelity as a function of the photon number for different stacks.
We find that cluster states on a smaller stack have lower fidelity than those made from larger stacks.
This observation is due to the increased number of photons built up in the third dimension, which increases the number of imperfect entanglements.
Furthermore, the fidelities of the $N\times M$ and $M\times N$ stacks are not equivalent.
This asymmetric fidelity is due to the helical boundary condition of the photonic cluster states (see Section~\ref{sec:two}).
However, we fit an exponential function to these results from the different stack sizes and find that all feature a scaling factor $\left(\beta\right)$ of $0.97$ per photon. Therefore, the stack dimension only affects the fidelity of the first few photons and does not affect the scaling behavior of the fidelity.

\begin{figure}[b]
\includegraphics[width=3.2in]{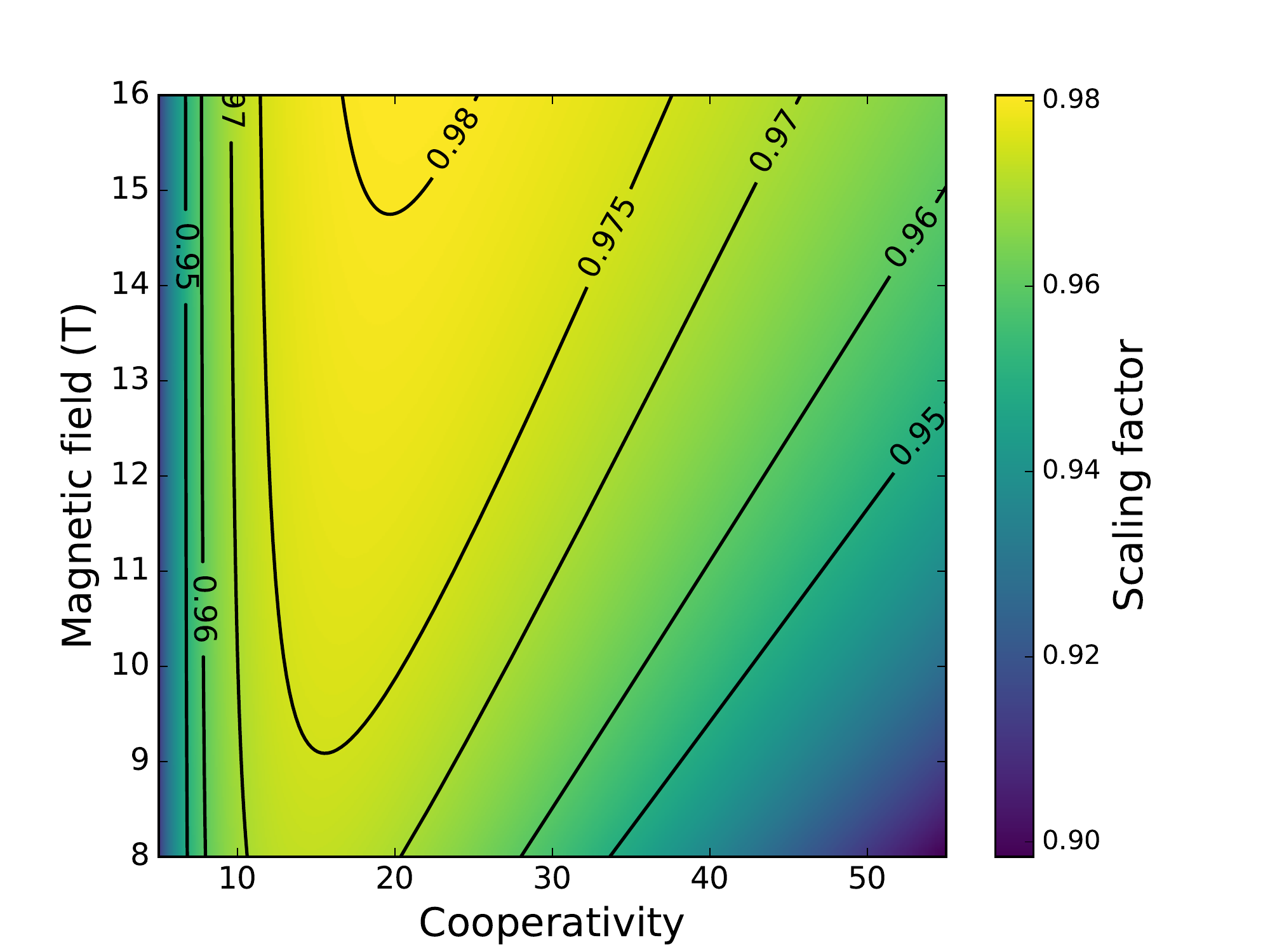}
\caption{\label{fig:10}
The scaling factor $\left(\beta\right)$ of the fidelity as a function of the resonant cooperativity and magnetic field for a 3D cluster state.}
\end{figure}

The cavity cooperativity determines the interaction strength between the photon and the quantum dot, impacting the fidelity of the output states.
Therefore, we investigate the fidelity under different resonant cooperativity $C_\uparrow$ and magnetic field $B$.
These two parameters uniquely determine the off-resonant cooperativity $C_\downarrow$.
Figure~\ref{fig:10} shows the scaling factor $\left(\beta\right)$ of the fidelity  as a function of the cooperativity $C_\uparrow$ and magnetic field $B$.  The scaling factor is calculated by curve fitting of the fidelity as a function of photon number for the 3D cluster states with the stack of $2\times2$.
The fidelity increases with the strength of the magnetic field.
Additionally, we find an optimal cooperativity for each magnetic field value to maximize the output state's fidelity.
This behavior is surprising given the prevailing assumption that higher cooperativity atom-cavity systems are inherently better for single-photon entanglement applications.
This counter-intuitive behavior arises because at a fixed magnetic field, if the resonant cooperativity $C_\uparrow$ is too low, the photon does not interact with the atom.
However, if $C_\uparrow$ is too high, the off-resonant $C_\downarrow$ cooperativity will also increase, according to the relation $C_\downarrow=\frac{C_\uparrow}{1+i2\Delta_z/\kappa}$, exceeding the requirement for a high-quality CPHASE gate $\left(C_\downarrow\ll1\right)$, thus decreasing the fidelity.

\section{Chiral coupling}
The previous analysis shows that high fidelity cluster states require extremely high magnetic fields since both spin-up and spin-down states couple to the same cavity mode.  However, it is impractical to apply such a high magnetic field because increasing the magnetic field will decrease the relaxing time of the spin in the quantum dot \cite{Lu2010}.
Here, we analyze a potentially better approach based on “chiral coupling” \cite{Lodahl2017}, as described below.
This approach eliminates the tradeoff, enabling high fidelities even with no Zeeman splitting.

\begin{figure}[t]
\subfloat[\label{fig:11a}]{}
\subfloat[\label{fig:11b}]{}
\subfloat[\label{fig:11c}]{}
\includegraphics[width=3.4in]{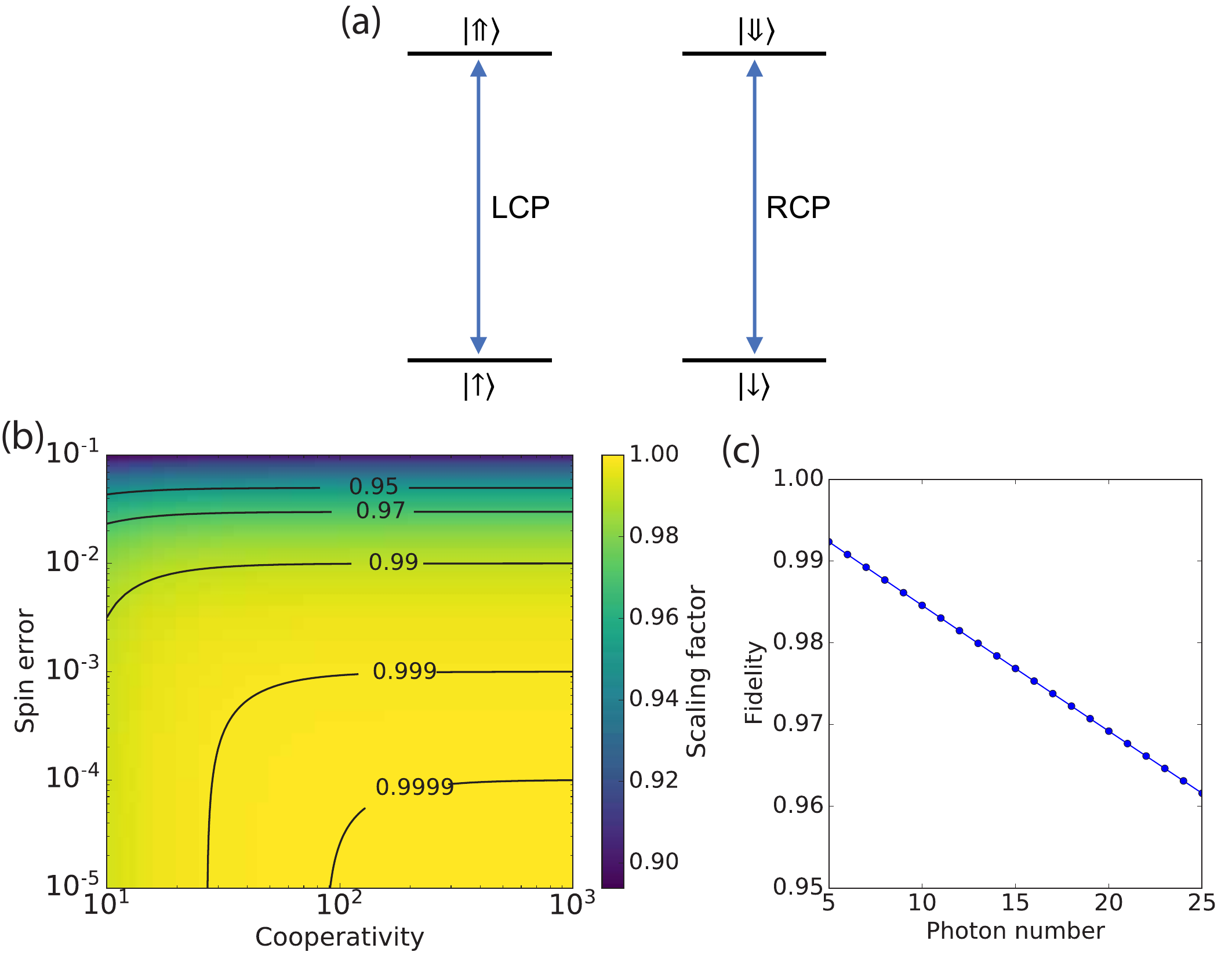}
\caption{\label{fig:11}
(a) The level structures of a cavity-coupled atomic qubit in chiral coupling. LCP/RCP means left/right-handed circularly polarized light.
(b) Contour plots showing the scaling factor $\left(\beta\right)$ of the fidelity  as a function of the cooperativity and spin error.
(c) The fidelity as a function of the photon number in cluster states with cooperativity $C=100$, spin rotation error $p=0.001$, and dephasing time $T_2=\SI{5}{\micro\second}$.}
\end{figure}

Figure~\ref{fig:11a} illustrates the specific atomic-level structure we consider.
The atomic qubit possesses two ground states $(\left|\uparrow\right\rangle, \left|\downarrow\right\rangle)$ and two excited states $(\left|\Uparrow\right\rangle, \left|\Downarrow\right\rangle)$, where the spin quantization axis is along the longitudinal direction of the cavity.
We assume the cavity has two degenerate modes whose electric field is circularly polarized at the location of the spin-qubit.
This condition creates the chiral coupling, where the spin-down state only couples to the left-handed circularly polarized mode while the spin-up state only couples to the right-handed circularly polarized mode \cite{Junge2013, Shomroni2014, Barik2020}.
Due to this selection rule, a reflected photon will acquire a phase depending on its polarization and the spin state.
The reflection coefficient of the cavity is given by (see Appendix~\ref{app:b})
\begin{equation}
    r=\frac{g^2-\frac{\gamma}{2}\left[i\Delta_s+\frac{\kappa}{2}\right]}{g^2+\frac{\gamma}{2}\left[i\Delta_s+\frac{\kappa}{2}\right]}\;,
\end{equation}
where $\Delta_s$ is the frequency detuning between the photon and atomic transition; $g$, $\kappa$, and $\gamma$ are the atom-cavity coupling strength, atom dipole decay rate, and cavity decay rate, respectively.
When the atom does not couple to the polarized photon $\left(g=0\right)$, the reflection coefficient $r_0$ equals $-1$.
When the atom couples to the polarized photon, the reflection coefficient strongly depends on the detuning.
We can choose $\Delta_s=\frac{\kappa}{2}\sqrt{C^2-1}$ to make the reflection with a phase shift of $\frac{\pi}{2}$.
The reflection coefficient is given by $r_1=-i\sqrt{\frac{C-1}{C+1}}$ where $C=\frac{4g^2}{\gamma\kappa}$ is the atom-cavity cooperativity.
In the high cooperativity limit $\left(C\gg1\right)$, the reflection coefficient $r_1=-i$.
We define the state of the incident photon as $\left|L\right\rangle=\left|0\right\rangle_p$ for a left circularly polarized photon and $\left|R\right\rangle=\left|1\right\rangle_p$ for a right circularly polarized photon.
We similarly label the spin states as  $\left|\uparrow\right\rangle\equiv{\left|0\right\rangle}_A$ and $\left|\downarrow\right\rangle\equiv{\left|1\right\rangle}_A$.
The state of the total system, denoted as $\left|\psi\right\rangle=|x\rangle_p\otimes |y\rangle_A$ where $x$ and $y$ are the qubit states of the photon and spin respectively, transforms according to the following operator
\begin{equation}
U_{CR}=
    \begin{pmatrix}
    r_1 & 0 & 0 & 0\\
    0 & -1 & 0 & 0\\
    0 & 0 & -1 & 0\\
    0 & 0 & 0 & r_1
    \end{pmatrix}\;.
\end{equation}
This transformation is not a CPHASE gate but can easily be transformed into one by applying phase shifts to the spin and photon given by the phase operator $\left(\begin{matrix}1&0\\0&i\\\end{matrix}\right)\otimes\left(\begin{matrix}1&0\\0&i\\\end{matrix}\right)$.
This CPHASE gate does not require large Zeeman splitting between the two optical transitions.
It only requires high cooperativity to achieve high fidelity.

We now analyze the fidelity of cluster states generated using chiral coupling.
Figure~\ref{fig:11b} plots the scaling factor $\left(\beta\right)$ of the fidelity for a 3D cluster state as a function of the cooperativity $C$ and spin error $1-\left(1-\frac{p}{2}\right)\left(1-\frac{\tau}{T_2}\right)$, which accounts for both imperfect spin rotation and dephasing as defined previously.
The fidelity increases with the cooperativity until an upbound determined by the spin error.
This is because the CPHASE gate is nearly perfect when the cooperativity is large, and the spin error will be the only factor affecting the fidelity of cluster states.
We can achieve a scaling factor of the fidelity of about $0.999$ with a spin error rate of $0.001$, which is above the threshold (an error rate of $0.75\%$) of fault-tolerant cluster state quantum computation \cite{Raussendorf2007}.
These spin error rates are achievable in both atomic \cite{Campbell2010, Harty2014, Gaebler2016} and solid-state qubits \cite{Yoneda2018}, but combining these error rates with high cooperativities remains a significant challenge.
Figure~\ref{fig:11c} plots the fidelity as a function of the number of photons in the cluster state with cooperativity $C=100$, spin rotation error $p=0.001$, and dephasing time $T_2=\SI{5}{\micro\second}$.
By fitting the data with an exponential function, we see a fidelity decay factor of $0.998$ per photon of the cluster states $\left(\mathcal{F}_1 \sim {0.998}^K\right)$.

\section{Conclusion and outlook}
We have proposed and analyzed a protocol that can efficiently generate multi-dimensional photonic cluster states using an atom-cavity system and a small number of time-delay feedback.
The protocol is deterministic under ideal conditions and therefore could achieve scalable photonic cluster states.
Our results highlight the importance of chiral coupling to achieve high fidelity without the need for extremely large Zeeman splitting that requires impractical magnetic fields.
A note added that a recent paper by Kianna et al. \cite{Wan2020} also proposed a protocol to generate three-dimensional cluster states using constant ancilla and delay lines. They gave a good analysis of the propagation of Pauli errors in the generated states. In comparison, our work demonstrates a physical implementation using an atom-cavity system. We also provide a formalism to bridge and analyze the experimental imperfections and the logical errors of the generated states.

The decomposing and contracting method of deriving tensor network states, as shown in Figures~\ref{fig:7} and~\ref{fig:8}, is also valuable for broader scenarios of generating multiqubit states.  For example, this method provides a diagrammatic derivation of matrix product states (MPS) sequentially generated by a single-photon source in a cavity \cite{Schon2005, Schon2007}.  One can also include errors by replacing the unitary operators with Kraus operators \cite{Verstraete2004a, Zwolak2004}.  Furthermore, our derivation has a particular advantage in simulating projected entangled pair states (PEPS) generated using delay lines.  In those systems, some qubits can interact multiple times, which implements the logical long-range interactions, such that a tensor network representation of the state is hard to calculate.  However, our method can solve it by decomposing the two-qubit operators into two “entangled” local operator tensors.  The tensor network state is calculated by contracting the operator tensors of one site.  A potential application of the method is the simulation of high-dimensional Boson sampling \cite{Lubasch2018, Deshpande2021}.

For future studies, we expect to apply the time-delay feedback method to generate other types of tensor network states such as the Toric Code \cite{Kitaev2003}.
Ultimately, our protocol provides a versatile way to engineer complex entanglement that could enable a variety of applications in measurement-based quantum computation, quantum error correction, and quantum communication.

\begin{acknowledgments}
The authors would like to acknowledge support from the Air Force Office of Scientific Research (Grant No. FA9550-16-1-0323, FA95501610421), the Physics Frontier Center at the Joint Quantum Institute, the National Science Foundation (Grant No. PHYS. 1915371, OMA1936314, ), and the Maryland-ARL Quantum Partnership.
\end{acknowledgments}

\appendix

\section{\label{app:a}Derivation of the quantum circuit for generating cluster states}
We first prove the quantum circuit in Figure~\ref{fig:2b} generates the nearest neighbor entanglement between photons in a linear chain (1D cluster state).
We start with a representation of 1D cluster state with $K+1$ qubits \cite{Briegel2001}
\begin{equation}
|\Psi_{1D}\rangle=\prod_{k=1}^{K}{Z_{k-1,k}{|+\rangle}_A{\otimes|+\rangle}^{\otimes K}}\;,
\label{eq:A1}
\end{equation}
where $Z_{i,j}$ is the CPHASE gate between qubit $i$ and $j$; the $0^{th}$ qubit represents the ancilla (A), and other numbered qubits represent photons.
We introduce a swap operator $S_{i,j}$ which exchanges the states between qubit $i$ and $j$.
Thus, Eq.~\ref{eq:A1} can be rewritten as
\begin{equation}
|\Psi_{1D}\rangle=\prod_{k=1}^{K}{S_{A,k}Z_{A,k}{|+\rangle}_A{\otimes|+\rangle}^{\otimes K}}\;.
\label{eq:A2}
\end{equation}
The product is ordered from right to left, i.e. $\prod_{i=1}^{K}U_i=U_K\cdots U_1$.
We can use the relation
\begin{equation}
S_{A,k}Z_{A,k}{|\phi\rangle}_A\otimes{|+\rangle}_k=\left(H_A\otimes H_k\right)Z_{A,k}{|\phi\rangle}_A\otimes{|+\rangle}_k\;,
\label{eq:A3}
\end{equation}
where ${|\phi\rangle}_A$ is an arbitrary state of the ancilla; $H_A$ and $H_k$ are Hadamard gates on the ancilla and the photon, respectively.
This relation holds when the photon is in state $|+\rangle$, but the ancilla can be in an arbitrary state, either unentangled or entangled.
Using the relation, we reformulate Eq.~\ref{eq:A2} as
\begin{equation}
|\Psi_{1D}\rangle=\prod_{k=1}^{K}{\left(H_A\otimes H_k\right)Z_{A,k}{|+\rangle}_A{\otimes|+\rangle}^{\otimes K}}\;,
\label{eq:A4}
\end{equation}
which is implemented by the quantum circuit in Figure~\ref{fig:2b}.

We now prove Figure~\ref{fig:2d} generates the non-nearest neighbor entanglement between photons $k-N$ and $k$.
Based on the one-dimensional cluster state, we can add one dimension by entangling photons $k-N$ and $k$ as
\begin{equation}
|\Psi_{2D}\rangle=\prod_{k=N+1}^{K}Z_{k-N,k}\prod_{k=1}^{K}{Z_{k-1,k}{|+\rangle}_A{\otimes|+\rangle}^{\otimes K}}\;.
\label{eq:A5}
\end{equation}
Because CPHASE gates $Z_{i,j}$ commute each other, we can reorder the operators in Eq.~\ref{eq:A5}
\begin{eqnarray}
|\Psi_{2D}\rangle=&&\prod_{k=N+1}^{K}{Z_{k-1,k}Z_{k-N,k}}\nonumber\\
& &\times
\prod_{k=1}^{N}{Z_{k-1,k}{|+\rangle}_A{\otimes|+\rangle}^{\otimes K}}\;.
\label{eq:A6}
\end{eqnarray}
By introducing the swap operator $S_{i,j}$ as in the one-dimensional case, we can rewrite Eq.~\ref{eq:A6} as
\begin{eqnarray}
|\Psi_{2D}\rangle=&&\prod_{k=N+1}^{K}{S_{A,k}Z_{A,k}Z_{A,k-N}}\nonumber\\
& &\times
\prod_{k=1}^{N}{S_{A,k}Z_{A,k}{|+\rangle}_A{\otimes|+\rangle}^{\otimes K}}\;.
\label{eq:A7}
\end{eqnarray}
After using the relation of Eq.~\ref{eq:A3}, we rewrite Eq.~\ref{eq:A7} as
\begin{eqnarray}
|\Psi_{2D}\rangle=&&\prod_{k=N+1}^{K}{\left(H_A\otimes H_k\right)Z_{A,k}Z_{A,k-N}}\nonumber\\
& &\times
\prod_{k=1}^{N}{\left(H_A\otimes H_k\right)Z_{A,k}{|+\rangle}_A{\otimes|+\rangle}^{\otimes K}}\;.
\label{eq:A8}
\end{eqnarray}
Therefore, we first generate a 1D cluster state.
Based on it, the quantum circuit in Figure~\ref{fig:2d} implements the iterative steps of $\prod_{k=N+1}^{K}{\left(H_A\otimes H_k\right)Z_{A,k}Z_{A,k-N}}$, which entangles photon $k$ with photons $k-N$ and $k-1$, to generate the 2D cluster state.

Then we prove that Figure~\ref{fig:3b} generates the unit cell of the 3D cluster state.
By implementing two non-nearest neighbor gates to the linear chain of photons, the 3D cluster state can be generated by
\begin{eqnarray}
|\Psi_{3D}\rangle=&&\prod_{k=MN+1}^{K}Z_{k-MN,k}\prod_{k=N+1}^{K}Z_{k-N,k}\nonumber\\
& &\times
\prod_{k=1}^{K}Z_{k-1,k}{|+\rangle}_A{\otimes|+\rangle}^{\otimes K}\;.
\label{eq:A9}
\end{eqnarray}
Because $Z_{i,j}$ commute with each other, we can reorder the operators in Eq.~\ref{eq:A9}
\begin{eqnarray}
|\Psi_{3D}\rangle=&&\prod_{k=MN+1}^{K}{Z_{k-1,k}Z_{k-N,k}Z_{k-MN,k}}\nonumber\\
& &\times
\prod_{k=N+1}^{MN}{Z_{k-1,k}Z_{k-N,k}}\nonumber\\
& &\times
\prod_{k=1}^{N}Z_{k-1,k}{|+\rangle}_A{\otimes|+\rangle}^{\otimes K}\;.
\label{eq:A10}
\end{eqnarray}
After introducing the swap operator $S_{i,j}$ between the qubit $i$ and $j$, we can rewrite Eq.~\ref{eq:A10} as
\begin{eqnarray}
|\Psi_{3D}\rangle=&&\prod_{k=MN+1}^{K}{{S_{A,k}Z_{A,k}Z_{A,k-N}Z_{A,k-MN}}}\nonumber\\
& &\times
\prod_{k=N+1}^{MN}{S_{A,k}Z_{A,k}Z_{A,k-N}}\nonumber\\
& &\times
\prod_{k=1}^{N}{S_{A,k}Z_{A,k}}{|+\rangle}_A{\otimes|+\rangle}^{\otimes K}\;.
\label{eq:A11}
\end{eqnarray}
The operators in the second and third parentheses generate the 2D cluster state with dimensions of $M\times N$.
With the relation of Eq.~\ref{eq:A3}, we can rewrite the operators in the first parenthesis as
\begin{eqnarray}
|\Psi_{3D}\rangle=&&\prod_{k=N+1}^{K}{{\left(H_A\otimes H_k\right)Z_{A,k}Z}_{A,k-N}Z_{A,k-MN}}\nonumber\\
& &\times
|\Psi_{{2D}_{M\times N}}\rangle\;.
\label{eq:A12}
\end{eqnarray}
The quantum circuit in Figure~\ref{fig:3b} implements the iterative steps of $\prod_{k=N+1}^{K}{{\left(H_A\otimes H_k\right)Z_{A,k}Z_{A,k-N}Z_{A,k-MN}}}$ in Eq.~\ref{eq:A12}, which generates the 3D cluster state based on a 2D cluster state.

\section{\label{app:b}Derivation of the reflection coefficient}

The Hamiltonian of the atom-cavity system in Figure~\ref{fig:4} is given by $H=H_0+V$, where
\[
H_0=\hbar\omega_c{\hat{a}}^\dag\hat{a}+\hbar\omega_s\sigma_{\Uparrow\Uparrow}+\hbar\left(\omega_s-\delta_1\right)\sigma_{\Downarrow\Downarrow}+\hbar\delta_2\sigma_{\downarrow\downarrow}
\]
\begin{eqnarray*}
V=i\hbar g_V\sigma_{\Uparrow\uparrow}\hat{a}+i\hbar g_V\sigma_{\Downarrow\downarrow}\hat{a}+i\hbar g_C\sigma_{\Downarrow\uparrow}\hat{a}+&&i\hbar\sigma_{\Uparrow\downarrow}\hat{a}\\
 & &+h.c.\;.
\end{eqnarray*}
We assume the cavity only support the vertically polarized mode, so $g_C=0$ and $g_V=g$.
The atomic state space can be reduced into two uncoupled subspaces $\left\{\left|\uparrow\right\rangle,\ \left|\Uparrow\right\rangle\right\}$ and $\left\{\left|\downarrow\right\rangle,\left|\Downarrow\right\rangle\right\}$.
In either subspace, the Heisenberg equations of motion for the cavity field operator $\hat{a}$, the atomic operator $\sigma_-$, and the external field input-output relation are given by \cite{Walls2008, Hu2008}
\begin{equation}
\begin{cases}
& \frac{d\hat{a}}{dt}=-\left[i\left(\omega_c-\omega\right)+\frac{\gamma}{2}\right]\hat{a}-g\sigma_--\sqrt{\gamma}{\hat{a}}_{in}\\
& \frac{d\sigma_-}{dt}=-\left[i\left(\omega_s-\omega\right)+\frac{\kappa}{2}\right]\sigma_--g\sigma_z\hat{a}\\
& {\hat{a}}_{out}={\hat{a}}_{in}+\sqrt{\gamma}\hat{a}
\end{cases}\;,
\label{eq:B1}
\end{equation}
where $\omega_c$, $\omega_s$, and $\omega$ are the frequencies of cavity field, atomic transition, and probe beam (external field); $g$, $\kappa$, and $\gamma$ are the atom-cavity coupling strength, atom dipole decay rate, and cavity decay rate, respectively.
We consider the singly-photon process, and the system operates in the linear weak excitation limit where $\sigma_z=-1$.
We also assume that the photon is quasi-monochromatic, which results in a steady-state reflection coefficient of
\begin{equation}
    r\left(\omega\right)=1-\frac{\gamma\left[i\left(\omega_s-\omega\right)+\frac{\kappa}{2}\right]}{\left[i\left(\omega_s-\omega\right)+\frac{\kappa}{2}\right]\left[i\left(\omega_c-\omega\right)+\frac{\gamma}{2}\right]+g^2}\;.
\label{eq:B2}
\end{equation}
We assume the cavity mode is on-resonant with the frequency of the external field $\left(\omega_c=\omega\right)$ and define the detuning $\Delta=\omega_s-\omega$, so the spin-dependent reflection coefficient is given by
\begin{equation}
    r_{\uparrow,\downarrow}=\frac{g^2-\frac{\gamma}{2}\left[i\Delta_{\uparrow,\downarrow}+\frac{\kappa}{2}\right]}{g^2+\frac{\gamma}{2}\left[i\Delta_{\uparrow,\downarrow}+\frac{\kappa}{2}\right]}\;,
\label{eq:B3}
\end{equation}
where $\Delta_\uparrow=0$, and $\Delta_\downarrow=\delta_1+\delta_2$ detunes by the Zeeman splitting.

For the system in Figure~\ref{fig:11a}, we assume the cavity has two degenerate modes whose electric field is circularly polarized at the location of the spin-qubit.
When the probe photon is in circular polarization, the total atomic state space can be reduced into two uncoupled state subspaces $\left\{\left|\uparrow\right\rangle,\left|\Uparrow\right\rangle\right\}$ and $\left\{\left|\downarrow\right\rangle,\left|\Downarrow\right\rangle\right\}$.
The reflection coefficient of the cavity can be calculated by considering either of the subspace.
When the spin is in the subspace of $\left\{\left|\uparrow\right\rangle,\left|\Uparrow\right\rangle\right\}$, the Hamiltonian is given by $H=H_0+V$, where
\begin{eqnarray*}
&H_0 &=\hbar\omega_c{\hat{a}}^\dag\hat{a}+\hbar\omega_s\sigma_{\Uparrow\Uparrow}\\
&V &=i\hbar g\sigma_{\Uparrow\uparrow}\hat{a}-i\hbar g\sigma_{\uparrow\Uparrow}{\hat{a}}^\dag\;.
\end{eqnarray*}
The operators’ equations of motion are the same as Eq.~\ref{eq:B1}, and we can get the reflection coefficient as
\begin{equation}
    r=\frac{g^2-\frac{\gamma}{2}\left[i\Delta_s+\frac{\kappa}{2}\right]}{g^2+\frac{\gamma}{2}\left[i\Delta_s+\frac{\kappa}{2}\right]}\;,
\end{equation}
where $\Delta_s$ is the frequency detuning between the photon and atomic transition; $g$, $\kappa$, and $\gamma$ are the atom-cavity coupling strength, atom dipole decay rate, and cavity decay rate, respectively.

\section{\label{app:c}Tensor decomposition}

In this appendix, we explain how to decompose the CPHASE gate.
We first write the matrix element of CPHASE gate between qubit $i$ and $j$ as
\begin{equation}
Z_{i,j}=Z_{\alpha^\prime,\beta^\prime}^{\alpha,\beta}\left|\alpha^\prime,\beta^\prime\rangle\langle\alpha,\beta\right|_{i,j}\;.
\end{equation}
Then we can partially transpose the matrix as
\begin{equation}
Z_{i,j}=Z_{\alpha^\prime,\alpha}^{\beta^\prime,\beta}\left|\alpha^\prime\rangle\langle\alpha\right|_i\otimes\left|\beta^\prime\rangle\langle\beta\right|_j\;.
\end{equation}
Then we can do a Singular Value Decomposition on the matrix $Z_{\alpha^\prime,\alpha}^{\beta^\prime,\beta}$on its row indices $\left(\alpha^\prime\alpha\right)$ and column indices $\left(\beta^\prime\beta\right)$ as
\begin{equation}
Z_{\alpha^\prime,\alpha}^{\beta^\prime,\beta}=\sum_{s}{U_{\alpha^\prime\alpha}^s\mathrm{\Lambda}_s{V_{\beta^\prime\beta}^s}^\ast}\;,
\end{equation}
where $U$ and $V$ are unitary, and $\Lambda$ is positive.  Then we can take the square root of $\Lambda$ and combine it into $U$ and $V$ as
\begin{eqnarray}
A_{\alpha^\prime\alpha}^s=U_{\alpha^\prime\alpha}^s{\sqrt\Lambda}_s\;,
\\
B_{\beta^\prime\beta}^s={V_{\beta^\prime\beta}^s}^\ast{\sqrt\Lambda}_s\;,
\end{eqnarray}
we use Einstein’s notation to omit summation operators.
So we can represent the CPHASE tensor $Z$ by the multiplication of two tensors $A$ and $B$ as
\begin{equation}
Z_{\alpha^\prime,\alpha}^{\beta^\prime,\beta}=A_{\alpha^\prime\alpha}^sB_{\beta^\prime\beta}^s\;.
\end{equation}
In Figure~\ref{fig:7b}, we draw tensor A as a yellow square and tensor B as a green square.
The $\alpha,\alpha^\prime$ and $\beta,\beta^\prime$ are indices represent the basis of the qubit, and $s$ are index representing the entanglement bond between qubits.

\section{\label{app:d}Derivation of the imperfect photon-photon CPHASE gate}

In this appendix, we show the imperfect reflection operation $U$ in Eq.~\ref{eq:2} will result in an effective imperfect photon-photon CPHASE gate.
The quantum circuit in Figure~\ref{fig:3b} entangles photon $k$ with its neighbors in three-dimensional cluster states.
We can represent the circuit in an operator form as
\begin{equation}
    \mathcal{E}=H_s \otimes H_k Z_{s,k}Z_{s,k-N}Z_{s,k-MN}\;.
    \label{eq:D1}
\end{equation}
In the real physical implementation, we should replace the ideal CPHASE gates $Z$ by the reflection operator $U_{s,k}=\left(I+\epsilon R_{s,k}\right)Z_{s,k}$, where $\epsilon R=UZ-I$ and $\epsilon \sim O\left(\left|1-\frac{C_\uparrow-1}{C_\uparrow+1}\right|+\left|1+\frac{C_\downarrow-1}{C_\downarrow+1}\right|\right)$ indicates the order of errors.
Thus, Eq.~\ref{eq:D1} transforms as
\begin{equation}
    \mathcal{E}_1 = \left(H_s \otimes H_k\right) \left(I+\epsilon R_{s,k}\right) Z_{s,k}U_{s,k-N}U_{s,k-MN}\;.
    \label{eq:D2}
\end{equation}
By commuting $H_s\otimes H_k$ and $I+\epsilon R_{s,k}$ then using the relation of Eq.~\ref{eq:A3}, we can simplify Eq.~\ref{eq:D2} as
\begin{equation}
    \mathcal{E}_1=\widetilde{U}_{s,k} U_{k,k-N} U_{k,k-MN} S_{s,k}\;,
    \label{eq:D3}
\end{equation}
where $S_{s,k}$ is the swap gate between the ancilla and photon $k$, and ${\widetilde{U}}_{s,k}=\left[I+\epsilon\left(H_s\otimes H_k\right)R_{s,k}\left(H_s\otimes H_k\right)\right]Z_{s,k}$ is also an imperfect CPHASE gate with the error in $O\left(\epsilon\right)$.
Finally, we achieve an effective operation for entangling photon $k$ with its six neighbors as
\begin{equation}
    \mathcal{E}=U_{k+MN,k} U_{k+N,k} U_{k+1,k} U_{k,k-N} U_{k,k-MN} U_{k,k-1}\;,
\end{equation}
where $\widetilde{U}$ denotes the gate between nearest neighbors, $U$ denotes the gate between non-nearest neighbors.


\bibliographystyle{apsrev4-2}
\bibliography{multi_dim_cluster_states}

\end{document}